\documentclass[aps,prl,twocolumn,groupedaddress]{revtex4}

\usepackage{graphicx}

\begin{document}

\renewcommand{\vec}[1]{{\mathbf #1}}

\title{Dynamics of Weakly Localized Waves}
\author{S.E. Skipetrov}
\email[]{Sergey.Skipetrov@grenoble.cnrs.fr}
\author{B.A. van Tiggelen}
\email[]{Bart.Van-Tiggelen@grenoble.cnrs.fr}
\affiliation{Laboratoire de Physique et Mod\'elisation des Milieux Condens\'es/CNRS,\\
Maison des Magist\`{e}res, Universit\'{e} Joseph Fourier, 38042
Grenoble, France}

\date{\today}

\begin{abstract}
We develop a transport theory to describe the dynamics of (weakly)
localized waves in a quasi-1D tube geometry both in reflection and
in transmission. We compare our results to recent experiments with
microwaves, and to other theories such as random matrix theory and
supersymmetric theory.

 \end{abstract}

\pacs{}

\maketitle

% ***********************************************************

Localization of waves has always been among the most difficult yet
most fascinating topics in the study of wave propagation in
disordered media. The first studies dealt with infinite media,
showing that localization is always achieved in 1D, but that a
minimum amount of disorder is required in dimensions larger than 2
\cite{anderson58}. In 3D the critical point is estimated by the
Ioffe-Regel criterion $k\ell \approx 1$, with $k$ the wavenumber
and $\ell$ the mean free path of the waves at a specified
frequency \cite{ping}. Later studies  \cite{loc} have considered
localization in open media, and emphasized the `leakage' through
the boundaries --- quantified by the conductance --- as the basic
localization parameter. The Thouless criterion \cite{thou} states
that `leaky', extended states become localized when the
`dimensionless conductance' $g=G/(2e^2/h)$ is of order one. Most
recent studies, both in theory \cite{carlo} and experiment
\cite{chabanov00}, have emphasized the giant fluctuations in
transmission coefficients in the regime $g<1$, confirming the
fundamental importance of the Thouless conductance for all
localization phenomena. In the diffuse regime ($g \gg 1$),
apart from a factor of order unity, the
dimensionless conductance $g$ can be expressed as the ratio of the
inverse microscopic level spacing, called the Heisenberg time $t_{H}$, and
the Thouless time $t_D = (L + 2 z_0)^2/\pi^{2} D_B$ (with $D_B$ the diffusion constant,
$L$ the size of the medium,
and $z_0 \sim \ell$ accounting for internal reflection).

A theory for `all localization' does not exist. Important elements
should be its capability to describe the transition from the
diffuse to the localized regime, notably with regard to leakage
and dynamics, and in all dimensions, and its flexibility to
experimental details, such as internal reflection, anisotropic
scattering and absorption. A very complete localization theory is
random matrix theory \cite{carlo}. It can describe the transition
from weak to strong localization, scaling, absorption, and
fluctuations, and recently also dynamics \cite{schomerus} but
applies only for low-dimensional systems. Supersymmetric theory
\cite{efetov97,mirlin00} has also a very general range of applicability but
does not always give the necessary physical insight to guide experiments. Finally, the
self-consistent theory for localization \cite{vw80} holds in all
dimensions, is able to describe critical behavior around the
mobility edge \cite{wegner}, and has a clear generalization for
dynamical problems. Its major disadvantages are that it applies only
to the field correlation function and not to higher moment
statistics, and its failure in the case of broken time-reversal invariance.
It is valid on length scales larger than the mean free
path, and --- in the time domain --- for times less than the
Heisenberg time.

Leakage effects can be studied from the  `leakage function' (LF)
 $P_{T, R}(\alpha)$ defined from the ensemble-averaged, time-dependent
transmission (reflection) $I_{T,R}(t)$ according to,
\begin{eqnarray}
I_{T, R}( t) = \int\limits_0^{\infty} \textrm{d} \alpha
\exp\left( -\alpha t \right) P_{T, R}(\alpha). \label{cut}
\end{eqnarray}
Supersymmetric theories \cite{muz95,mirlin00} have predicted
strongly non-exponential decay in transmission, even for weakly
localized waves ($g \gg 1$) and in quasi-1D typically of the kind
`$\exp[-g \ln^{2} (t/t_H)]$' beyond the Heisenberg time. It is in
this regime that a modal picture is appropriate, like in chaotic
cavities \cite{weaver},  and that $P_{T,R}(\alpha)$ can be argued to
equal the genuine distribution of resonant widths $P(\Gamma)$ of
the modes \cite{pgamma} at small $\Gamma$. $P(\Gamma)$ has  a
log-normal behavior at very small $\Gamma$ attributed to
`prelocalized' modes \cite{mirlin00}, which have become a central
issue in the study of random lasers \cite{patra}.

For times smaller than the Heisenberg time $t_{H}$ supersymmetric
theory  predicts the transmission to decay like
$\exp[-t/t_D + (1/g \pi^{2}) t^{2}/t^2_{D}]$ \cite{mirlin00}.
This would imply a narrow Gaussian distribution for
$P_T(\alpha)$, centered around the average Thouless leakage
$1/t_D$ with width $\sim 1/\sqrt{g}$. A recent numerical simulation of
wave dynamics in 2D disordered media \cite{haney03} has shown a
similar, roughly quadratic increase of the logarithm of intensity.
Chabanov, Zhang, and Genack \cite{andrey03} recently studied weakly
localized microwaves in quasi-1D at times scales up to the
Heisenberg time, and observed a non-exponential transmission with
time of the same type.  Another interesting report --- coming from
random matrix theory \cite{titov00}, and first reported for purely
1D systems \cite{papa} --- is the $1/t^2$ reflection coefficient for
the semi-infinite quasi-1D tube, rather than
the familiar $1/t^{3/2}$ decay expected from diffusion theory.
This implies that in reflection $P_R(\alpha)\propto {\alpha}$ for
small $\alpha$.

Transport theory ought to be valid for times less than the
Heisenberg time, beyond which a modal picture takes over. The
recent developments in theory and experiment call for a transport
theory for the dynamics of (weakly) localized waves, and notably
for the leakage functions $P_{T,R}(\alpha)$ defined in
Eq.~(\ref{cut}). This is the subject of the present Letter. We
will show that these functions are broadened by interference
effects in a way compatible with observations and supersymmetric
theory. We  emphasize that for $\alpha$ larger than the inverse Heisenberg
time (the typical level spacing)
the equivalence between the leakage function $P_{T,R}(\alpha)$ and
the resonant width distribution $P(\Gamma)$ is not established,
and that $P_{T,R}(\alpha)$ sometimes takes negative values.

Constructive interferences can be included into transport theory
using the self-consistent theory of localization. In finite, open
media this requires the appearance of a dynamical, spatially
dependent diffusion constant $D(\vec{r},\Omega)$ \cite{bart00},
which can explain the observed rounding of coherent backscattering
of light near the mobility edge \cite{schuur00,bart00}, as well as
the non-Ohmic transmission \cite{wiersma97}. We will here study
the dynamics. Given a short release of energy at the source at
$t=0$, the central observable is the flux of ensemble-averaged photon
energy $I(\vec{r},t)$ at position $\vec{r}$ and at time $t$,
with Fourier transform $I(\vec{r}, \Omega)$ that we shall
continue analytically in the whole complex plane. By causality is
$I(\vec{r}, \Omega)$ an analytic function in the upper complex
sheet $\textrm{Im}\ \Omega > 0$. For positive times we can change
the contour of the inverse Fourier transform with respect to
frequency into the negative complex plane. If we assume that
simple poles or branch cuts appear only along the negative
imaginary axis, we find relation~(\ref{cut}) with
\begin{eqnarray}
P_{T,R}(\alpha) &=& -i \lim\limits_{\epsilon \downarrow 0} \left[
I_{T,R}\left(\Omega =  - i \alpha + \epsilon\right)
\right.
\nonumber \\
&-& \left. I_{T,R}\left(\Omega = - i \alpha -\epsilon
\right) \right]. 
\label{pa} 
\end{eqnarray}
In the normal
diffuse regime only simple poles show up at $\Omega_n= -i n^2/t_D$,
and $P_{T, R}(\alpha)$ equals an infinite sum of Dirac delta
distributions. Purely localized modes would show up as a
contribution $\delta(\alpha)$ at zero leakage, but occur only in
infinite or closed media. For an open quasi-1D system ($N \gg 1$
transverse modes, length $L \gg \ell$, classical diffusion constant $D_B = v_E \ell/3$,
transport mean free path $\ell \gg$ wavelength, and the energy transport velocity $v_E$)
the basic equation is the 1D dynamic diffusion equation for the
intensity Green's function $C(z,z^{\, \prime},\Omega)$,
\begin{eqnarray}
\left[ -i \Omega -  \partial_z D(z, \Omega) \partial_z \right]
C(z, z^{\, \prime}, \Omega) =  \delta(z - z^{\, \prime}), \label{difeq}
\end{eqnarray}
supplied by the self-consistency condition for the dynamic
diffusivity imposed by reciprocity \cite{bart00},
\begin{eqnarray}
\frac{1}{D(z, \Omega)} = \frac{1}{D_B} +  \frac{2}{\xi}\ C(z, z,
\Omega) , \label{selfcon}
\end{eqnarray}
featuring the length scale $\xi=\frac{2}{3}N\ell$. At the
boundaries $z=0, L$ we impose the usual radiative boundary
conditions $ C \mp z_0 [D(0/L,\Omega)/D_B]
\partial_z C = 0$,
where $z_0 \sim \ell$ accounts for internal reflection.
$I_{T,R}$ is related to $C$ through
$I_{T,R}(\Omega) = \mp D(z=L/0, \Omega)
\partial_z C(z=L/0, z^{\, \prime} = \ell, \Omega)$.

The stationary problem ($\Omega =0$) can be solved analytically by
the substitution $\textrm{d} \tau = \textrm{d} z /D(z, 0)$. This
shows that for $L \gg \xi$ the average transmission decays as $\exp(-L/\xi)$ which
identifies $\xi$ as the localization length. The diffuse regime
$L \ll \xi$ has normal Ohmic transmission with conductance $g \simeq g_0 =
\frac{4}{3}N\ell/(L + 2 z_0) \simeq 2\xi/L$. These results basically agree with
the ones obtained from the DMPK equation \cite{carlo} and
supersymmetric theory \cite{zirn}. Note that when $L \gtrsim \xi$
it is important to discriminate between $g_0$ and the real conductance $g$ which can be
much smaller by localization effects.

\begin{figure}
\includegraphics[width=0.47\textwidth,angle=180]{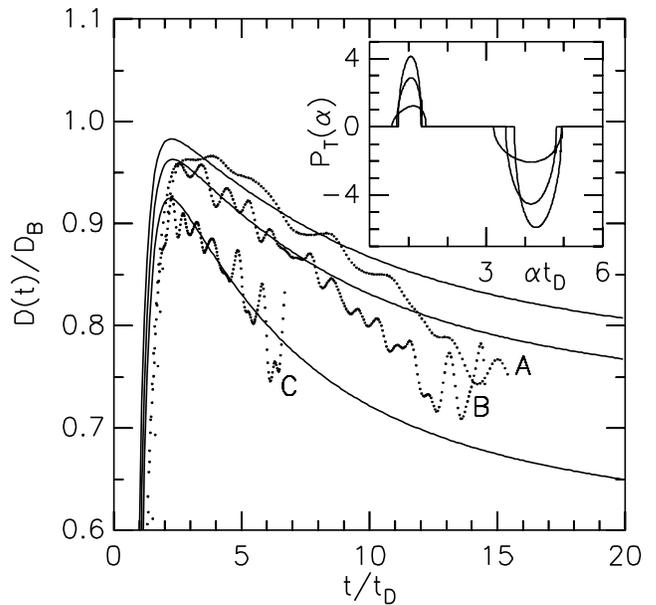}
\caption{\label{fig1} Time-dependent diffusion constant
for wave transmission through a quasi-1D disordered waveguide. Theoretical
results (solid lines) are compared to experimental data of Ref.\
\cite{andrey03} (dots). Satisfactory agreement between theory and
experiment for times $t$ below the Heisenberg time $t_{H}
\sim g_0 t_{D}$ is obtained by choosing $g_0$ equal to $9$
(sample A), 7.5 (B), and 4 (C) which is 14\% to 33\% larger than
the experimentally reported values. Inset: The leakage function
$P_T(\alpha)$ used to obtain the main plot.}
\end{figure}

The solution for any complex-valued  $\Omega$ has to be found
numerically, by iteration. For $g_0  \gtrsim 0.1$, we found satisfying and unique
convergence for all $\Omega$ after $10$--$100$ iterations.
We have evaluated the leakage function
by solving Eqs.\ (\ref{pa}--\ref{selfcon}) for $\epsilon = 10^{-9}/t_D$ and  $\epsilon = 10^{-10}/t_D$
and then using linear extrapolation to find the limit $\epsilon \downarrow 0$.
We have also carefully checked the
absence of singularities away from the negative imaginary axis.
The time-dependent transmission $I_T(t)$ was then obtained from
Eq.~(\ref{cut}). Absorption can be added but this will just give
rise to a trivial translation of the LF $P_T(\alpha)$ to higher
values for $\alpha$. Following Chabanov, Zhang and Genack \cite{andrey03}
we shall \emph{interpret} any non-exponential decay in terms of a
\emph{time-dependent} diffusion constant, in which case the
transmission would decay as
\begin{equation}\label{Dtt}
I_T(t) \sim \exp\left\{-\frac{\pi^{2}}{(L + 2 z_0)^{2}}\int_0^t \textrm{d}t^{\prime}\, D(t^{\prime})\right\}.
\end{equation}

In Fig.\ \ref{fig1} we have compared our calculation for $D(t)$
to experimental results obtained for three different choices for
the dimensionless conductance, corresponding to the samples $A$--$C$ of
Ref.\ \cite{andrey03}: $g_0 = 9$ (A), $g_0 = 7.5$ (B), and $g_0 = 4$ (C).
These values are slightly larger than can be estimated from the data of Ref.\
\cite{andrey03}. Our transport theory describes the experimental results
fairly well for all times below the Heisenberg time
$t_{H} \sim g_0 t_{D}$. The inset of Fig.\ \ref{fig1}
shows that the different branches of the leakage function  $P_T(\alpha)$
achieve a finite width, though all with finite support.
Note that the second branch has a negative value. We have fitted
the first, positive branch to a Gaussian distribution with the same
average and the same variance, and studied their variation with $g_0$.
The Gaussian distribution leads to a linear
decrease of $D(t)$  shortly after the diffusion time $t_D$. Our
findings can be summarized by the relation,
\begin{equation}\label{Dtlin}
\frac{D(t)}{D_{B}}  = 1 + \frac{A}{g_0}- \frac{B}{g_0}
\frac{t}{t_{D}},
\end{equation}
with $A=0.15$ and $B=0.20$. Supersymmetric theory \cite{mirlin00}
gives $B=2/\pi^2$ for orthogonal symmetry, in good agreement with
our value, but makes no report of $A$. Yet, we have noticed that
this term increases the agreement with experiment considerably.
Our theory assigns no weight to $P_T(\alpha)$ for values smaller
than a certain threshold $\alpha^* \approx
(1/t_D)(1-0.8/\sqrt{g_0})$, in strong disagreement with
supersymmetric theory \cite{muz95,mirlin00}, which predicts a
log-normal distribution for small $\alpha$, caused by
`prelocalized' states that have localization lengths much smaller than
the average localization length $\xi$.  Our transport theory is not
valid when $\alpha$ is small compared to the average level
spacing.
It is for this reason also that it can describe only the uninteresting
short-time dynamics of transmission in the localized regime $ g<1$ where the
Heisenberg time is smaller than the diffusion time.

\begin{figure}
\includegraphics[width=0.47\textwidth,angle=180]{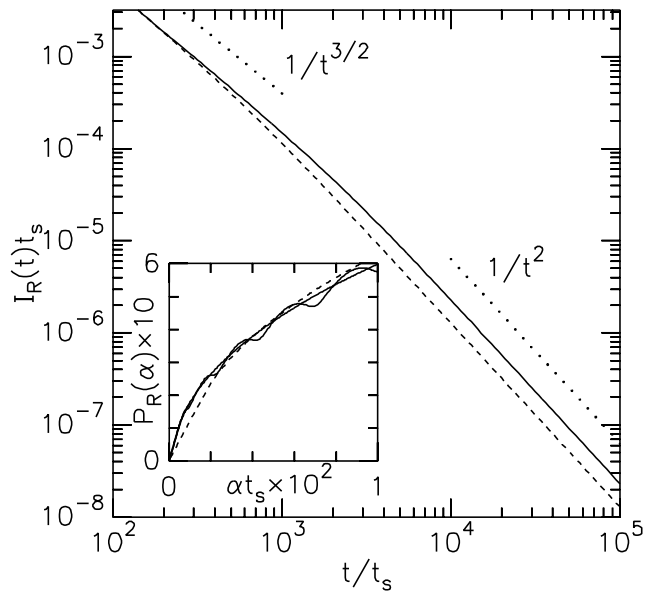}
\caption{\label{fig2} Time-dependent reflection from a quasi-1D
waveguide of length $L \gg \xi$: $N = 20$ (solid line) and $N = 10$ (dashed
line). Dotted lines show the slopes $1/t^{3/2}$ and $1/t^2$. Time has been
normalized by the mean free time $t_s$.
The curves are obtained by Laplace-transforming $P_R(\alpha)$ shown in the inset
($N = 20$, $L = 100 \ell$ --- wavy solid line;
$N = 20$, $L \rightarrow \infty$ --- solid line;
$N = 10$, $L \rightarrow \infty$ --- dashed line).}
\end{figure}

We will finally study the dynamics in reflection, and apply the
same procedure to calculate the leakage function $P_R(\alpha)$.
For $g \gg 1$ we find a series of clearly separated
branches, all \emph{positive} in sharp contrast to transmission,
and again with width $\sim 1/\sqrt{g}$.
Their maxima typically vary as $\sqrt{\alpha}$ which generates
the typically diffuse $1/t^{3/2}$ tail in the time domain.
The threshold leakage rate $\alpha^* \sim 1/t_D$ causes an exponential
decay at times beyond the diffusion time $t_D$.

As $g$ decreases the different
branches of $P_R(\alpha)$ start to join when $g_0 \approx 0.5$.
For $g_0 \ll 1$, the threshold leakage rate
decreases exponentially with $g_0$: $\ln \alpha^* \sim -1/g_0$, and
becomes rapidly very small, implying the disappearance
of exponential decay.
We will consider a waveguide of length $L \gg \xi$. We find that
when $L \gtrsim 20 \xi$, $P_R(\alpha)$ has converged to its
asymptotic limit at $L \rightarrow \infty$.
In this limit, $g_0 = 0$ and $t_H = \infty$, and our theory applies
at all times.
The asymptotic
$P_R(\alpha)$ roughly has a square-root behavior
that is taken over by
a \emph{linear} slope for small values of $\alpha$ (see the inset of
Fig.\ \ref{fig2}).
The linear law gives rise to the tail $I_R(t) \sim 1/t^2$ in the time-domain, as
can be seen in Fig.\ \ref{fig2}. This is
consistent with the prediction of Titov and Beenakker  using
random matrix theory \cite{titov00}. They have estimated the
cross-over to occur at a time $t \sim N^2 t_s$ (where $t_s$ is the mean free time),
again consistent with our findings.
We conclude that  this interesting dynamical cross-over is well captured by the
self-consistent transport theory, which, in contrast to the method
of Ref.\ \cite{titov00}, is not limited to the case of
$L \gg \xi$ and can be applied to a waveguide of any length
and at any time below the Heisenberg time $t_H$.

In conclusion, we have shown that the dynamics of (weak)
localization  both in transmission and in reflection of a quasi-1D
waveguide can be described by a self-consistent diffusion
equation. This theory is not valid beyond the Heisenberg time, and
other methods such as proposed by supersymmetric $\sigma$-models
have to be employed. Stimulated by recent accurate time-resolved
experiments of strongly disordered  3D materials close to the
mobility edge \cite{ad}, a future challenge is the application of
this theory to 2D and 3D systems.

We thank Felipe Pinheiro and Roger Maynard for many helpful
discussions, Andrey Chabanov and Azriel Genack for making
available their experimental results, and the GDR 2253 IMCODE of
the CNRS for support.

% ***********************************************************

\end{document}